\documentclass{webofc}
\usepackage[varg]{txfonts}   
\begin{document}
\title{Heavy-flavor impact on CTEQ-TEA global QCD analyses}
%
%

\author{\firstname{Marco} \lastname{Guzzi}\inst{1} \fnsep\thanks{\email{mguzzi@kennesaw.edu, presenter.}} 
          \and
        \firstname{Alim} \lastname{Ablat}\inst{2}
        \and
        \firstname{Sayipjamal} \lastname{Dulat}\inst{2}
        \and
        \firstname{Tie-Jiun} \lastname{Hou}\inst{3} 
        \and
        \firstname{Pavel} \lastname{Nadolsky}\inst{4}
        \and
        \firstname{Ibrahim} \lastname{Sitiwaldi}\inst{2}
        \and
        \firstname{Keping} \lastname{Xie}\inst{5}
        \and
        \firstname{C.-P.} \lastname{Yuan}\inst{6}
}

\institute{Department of Physics, Kennesaw State University, 370 Paulding Ave.,  30144 Kennesaw, GA, U.S.A.
\and
School of Physics Science and Technology, Xinjiang University, Urumqi, Xinjiang 830046 China
\and
School of Nuclear Science and Technology, Univ. of South China, Hengyang, Hunan 421001, China. 
\and 
Department of Physics, Southern Methodist University, Dallas, TX 75275-0181, U.S.A.
\and
Pittsburgh Particle Physics, Astrophysics, and Cosmology Center, Department of Physics and Astronomy, University of Pittsburgh, Pittsburgh, PA 15260, U.S.A.
\and
Department of Physics and Astronomy, Michigan State University, East Lansing, MI 48824 U.S.A.
          }

\abstract{%
We discuss heavy-flavor production at hadron colliders in recent global QCD analyses to determine parton distribution functions (PDFs) in the proton.
We discuss heavy-flavor treatments in precision theory predictions at the LHC. 
In particular, we discuss factorization schemes in presence of heavy flavors in proton-proton collisions, 
as well as the impact of heavy-flavor production at the LHC on PDFs. 
We show results of recent updates beyond CT18, the latest global QCD analysis from the CTEQ-TEA group.
}
\maketitle
\section{Introduction}
\label{intro}
Precise determination of parton distribution functions (PDFs) in the proton is critical for all current and future precision programs at the LHC.
Proton PDFs are an essential ingredient of QCD factorization theorems and are therefore ubiquitous in theory predictions for standard candle observables in hadronic collisions at high perturbative order in QCD. Precise and accurate theory predictions for such observables are necessary to investigate properties of the Higgs boson and to explore the Electroweak (EW) sector of the Standard Model (SM). In addition, they are important to scrutinize and validate SM extensions and search for signals of new physics interactions. 
PDFs are obtained through global QCD analyses of experimental hadronic cross section measurements by using a variety of analytical and statistical 
methods~\cite{Hou:2019efy,Bailey:2020ooq,NNPDF:2021njg,Alekhin:2017kpj}. 
They are ``data-driven'' quantities and they currently challenge precision in theory predictions at the LHC.         
In particular, precision measurements of heavy-quark (HQ) production observables in hadronic reactions play 
a significant role in PDF determinations because they can constrain PDFs in kinematic regions at intermediate and large parton momentum fraction $x$ which are currently poorly constrained by data. In addition, they provide us with complementary information besides that obtained from jet production in global QCD analyses.   
HQ production at the LHC at small transverse momentum $p_T$ and large rapidity $y$ of the HQ is sensitive to PDFs at both small and large $x$, where $x\approx \left( {\sqrt{p_T^2 +m_Q^2}}/\sqrt{S} \right) e^{\pm y}$. This allows us to probe QCD dynamics simultaneously in these two regimes. This is especially true for charm and bottom ($c/b$) quark production. In fact, in the $4 < |y| < 4.5 $ rapidity range at the LHC with a proton-proton collision energy of 13 TeV, it can probe $x\leq 10^{-5}$. 
When $p_T\geq 40$ GeV, it can probe $x\geq 0.2$.
On the other hand, top-quark pair production at the LHC, at the same collision energy, can probe the gluon PDF already at $x \gtrsim 0.01$, 
while $c/b$ production at HERA is sensitive to intermediate and small $x\lesssim 10^{-5}$. 
Probing these regimes (and beyond, at future facilities like the Electron Ion Collider (EIC)~\cite{Accardi:2012qut,Aschenauer:2017jsk,AbdulKhalek:2021gbh}, the Future Circular Collider (FCC)~\cite{FCC:2018byv,FCC:2018evy,FCC:2018vvp}, the Forward Physics Facility at the High-Luminosity (HL) LHC~\cite{Anchordoqui:2021ghd, Feng:2022inv}, and the Super proton-proton Collider (SppC)~\cite{Tang:2015qga}) will allow us to explore QCD factorization with an unprecedented level of accuracy. This 
will shed light on open questions like the intrinsic heavy-flavor content in the proton, and small-$x$ dynamics.
In this conference proceeding contribution, we shall report preliminary results of global PDF analyses which use recent high-precision HQ 
measurements from the LHC and HERA which are added on top of the CT18~\cite{Hou:2019efy} baseline. 
Moreover, treatments of QCD factorization in presence of HQ multiscale processes are also discussed.

\section{HQ treatments in QCD factorization and the S-ACOT-MPS scheme}
\label{sacot-mps}
\begin{figure}[h]
\centering
\includegraphics[width=6cm,clip]{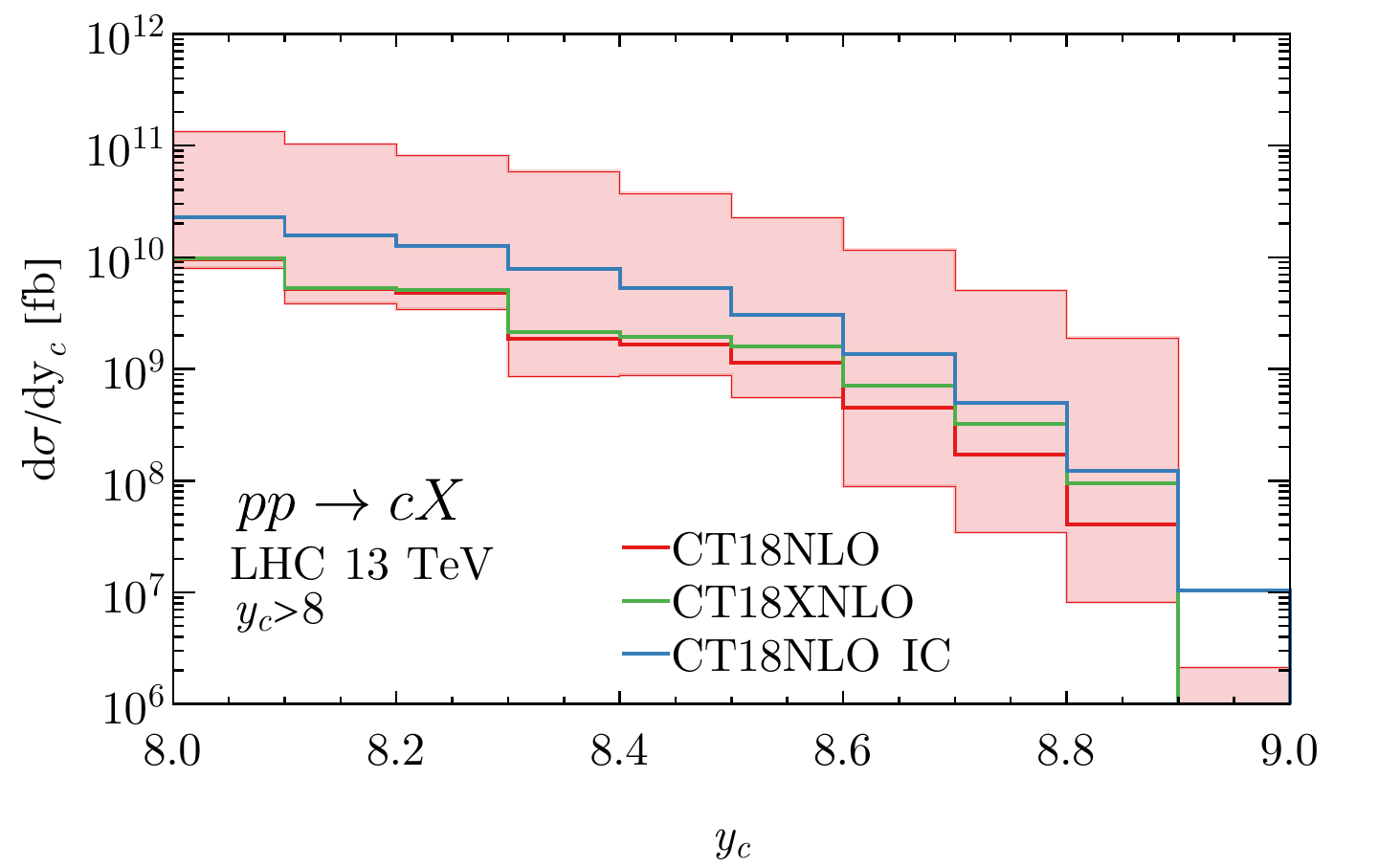}
\includegraphics[width=6cm,clip]{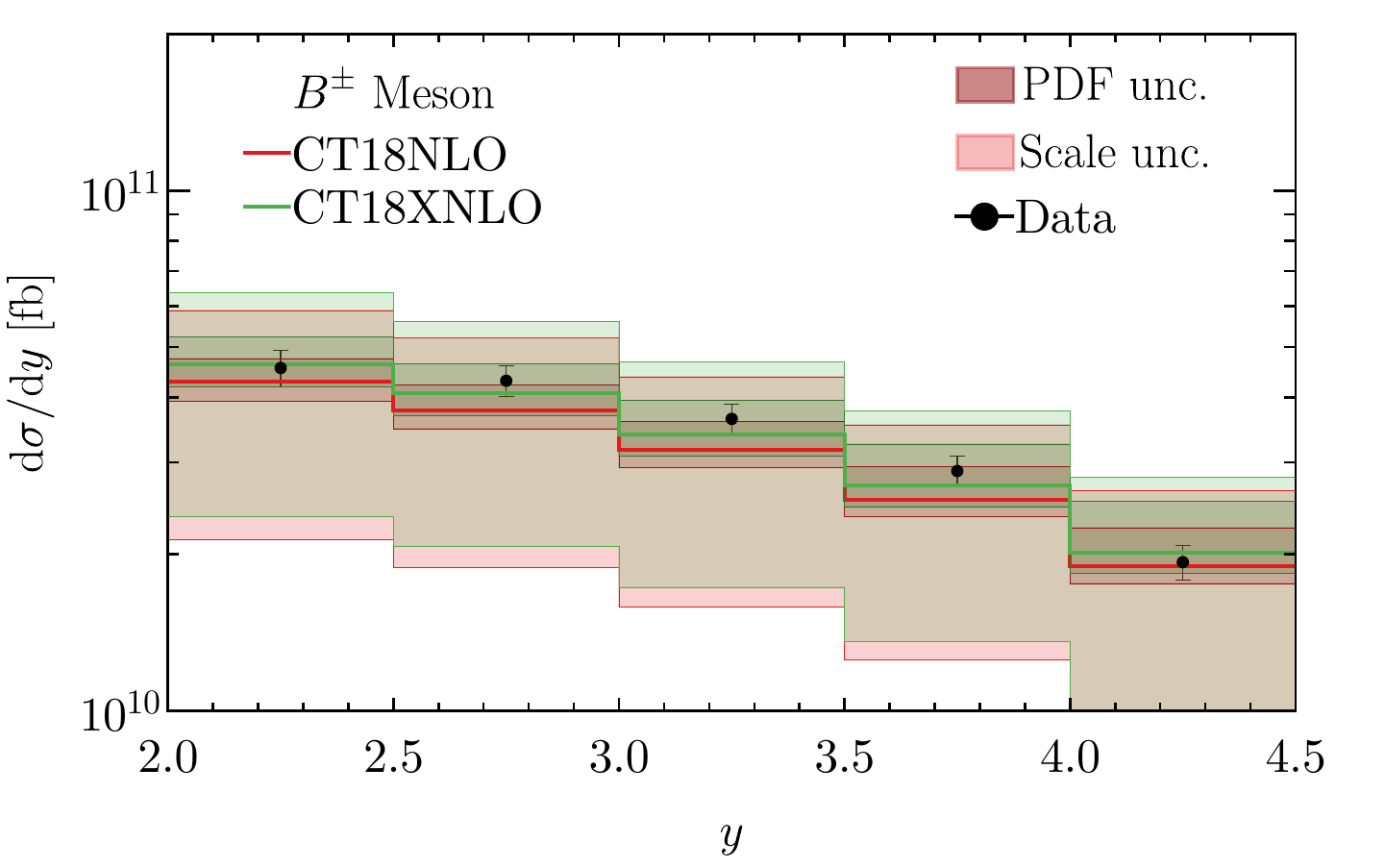}
\caption{{\bf Left}: Rapidity distributions of prompt charm at the LHC 13 TeV in the very forward region ($yc > 8$)~\cite{Anchordoqui:2021ghd}.
The error band represents the CT18NLO induced PDF uncertainty at 68\% CL. \\ {\bf Right}: NLO theory predictions for the rapidity distributions obtained with CT18NLO and CT18XNLO PDFs compared to $B^{\pm}$ production data~\cite{LHCb:2017vec} from LHCb 13 TeV.}
\label{SACOT-MPS}        
\end{figure}

HQ production dynamics is nontrivial due to the interplay of massless and massive schemes, which essentially are different ways of organizing the perturbation series.
In general, one can calculate cross sections by using massive schemes, that include HQs only in the final-state and apply at masses such that $p_T\lesssim m_Q$. In this case, the HQ $p_T$-spectrum can be obtained by keeping the number of flavors as fixed, {\it i.e.}, using a fixed-flavor number (FFN) scheme. That is, there is no HQ PDF in the proton; HQs are generated as massive final states, and $m_Q$ acts as an infrared cutoff. 
Moreover, all power terms of the type $\left(p_T^2/m_Q^2\right)^p$, where $p$ is a positive integer, are correctly accounted for in the perturbative series.
Cross sections can also be computed by using massless schemes when $p_T \gg m_Q \gg m_P$, where $m_P$ is the mass of the proton. 
In this case, large logarithmic terms of the type $\log^n \left(p_T^2/m_Q^2\right)$ spoil the convergence of the fixed-order expansion. Massless schemes are also known as zero-mass (ZM) schemes; the HQ is considered essentially massless everywhere and also enters the running of the strong coupling $\alpha_s$.
These large logarithms need to be resummed by using DGLAP evolution, that is, initial-state logs are resummed into a HQ PDF, while 
final-state logarithms are resummed into a fragmentation function (FF).
Modern global QCD analyses determine proton PDFs using a variety of experimental data which span wide areas of the $Q$-$x$ kinematic plane.
Therefore, amendments to the factorization formula are necessary to calculate key cross sections across wide ranges 
of energy and momentum transfer.   
Interpolating schemes, such as General Mass Variable Flavor Number (GMVFN) schemes, introduce modifications in QCD collinear factorization 
so that they appear as composite schemes which retain key mass dependence and efficiently resum collinear logarithms in order to combine the FFN and ZM schemes together. 
They are crucial for a correct treatment of HQ in deep inelastic scattering (DIS) processes~\cite{Aivazis:1993kh,Aivazis:1993pi,Buza:1996wv,Thorne:1997ga,Kramer:2000hn,Tung:2001mv,Alekhin:2009ni,Forte:2010ta,Guzzi:2011ew}
 and proton-proton ($pp$) collisions, and for accurate predictions of key scattering rates at the LHC in global QCD analyses to determine proton PDFs.
Moreover, they provide us with the possibility of directly accessing HQ PDFs parametrized at the initial scale.
This motivated the development of a general-mass (GM) factorization scheme for $pp$ collisions~\cite{Xie:2019eoe, Xie:2021ycd}, the Simplified ACOT scheme with Massive Phase Space (S-ACOT-MPS), which is based on an amended version of the S-ACOT scheme, developed for DIS~\cite{Aivazis:1993kh,Aivazis:1993pi,Kramer:2000hn,Tung:2001mv,Guzzi:2011ew}.  
Similar GMVFN schemes are studied in~\cite{Helenius:2018uul,Kniehl:2004fy,Kniehl:2005mk} and differences are related to the treatment of the phase space. 
The main idea behind a GMVFN scheme such as S-ACOT-MPS is the introduction of subtraction terms order-by-order in perturbation theory which avoid 
double counting and cancel enhanced collinear contributions from flavor-creation (FC) terms when ${\hat s} \gg m_Q^2$, or $p_T\gg M_Q$. 
In S-ACOT-MPS, the subtraction and flavor excitation (FE) contributions are evaluated in one single step. This improves stability in calculating differential cross sections and 
facilitates the extension of this scheme to higher orders.     
We applied S-ACOT-MPS to prompt $c$ and $b$ production at NLO in QCD at the LHC, at $\sqrt{S}=13$ TeV.   
In Figure~\ref{SACOT-MPS} (left), we illustrate the $c$-rapidity distribution determined with the CT18 and CT18X PDFs at NLO. We explore the sensitivity 
of prompt $c$ production to nonperturbative charm contributions in the proton by considering theory predictions in very forward regions with $y_c>8$.
This is relevant for future applications at the Forward Physics Facility at CERN~\cite{Anchordoqui:2021ghd} which will be able to 
clarify multiple aspects of QCD in such an extended forward region, in coordination with the HL-LHC and EIC.
In Figure~\ref{SACOT-MPS} (right), we compare theory predictions for the rapidity distribution of $B^\pm$ meson production at NLO in QCD 
with measurements~\cite{LHCb:2017vec} from the LHCb collaboration at $\sqrt{S}=13$ TeV. 
Theory uncertainties at NLO are large (${\cal O}(50\%)$) and mainly ascribed to scale variation. This can be improved by including 
higher-order corrections which require an extension of the S-ACOT-MPS scheme to NNLO.

\section{Impact on PDFs from high-precision $t {\bar t}$ production measurements}
\label{ttb-13TeV}
\begin{figure}[htb!]
\centering
\includegraphics[width=5.5cm,clip]{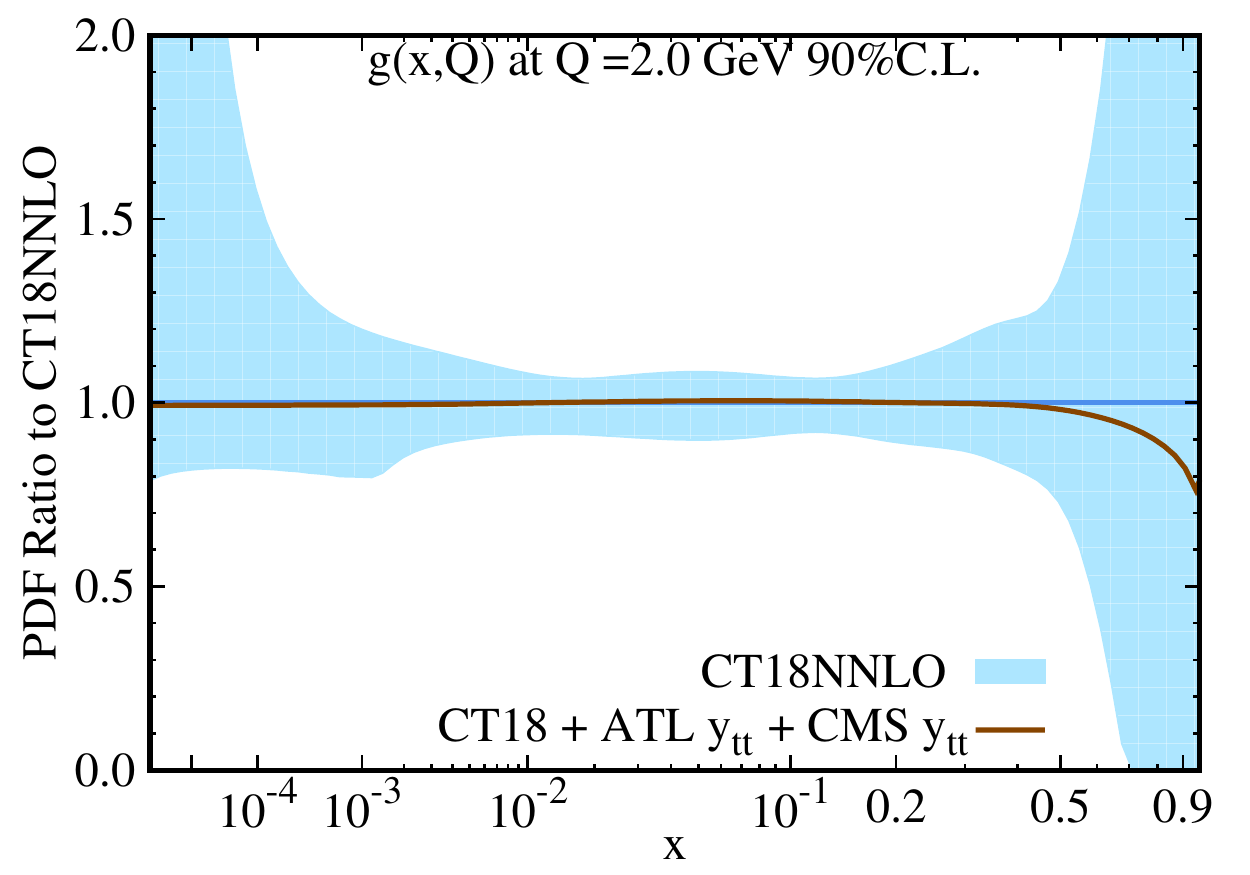}
\includegraphics[width=5.5cm,clip]{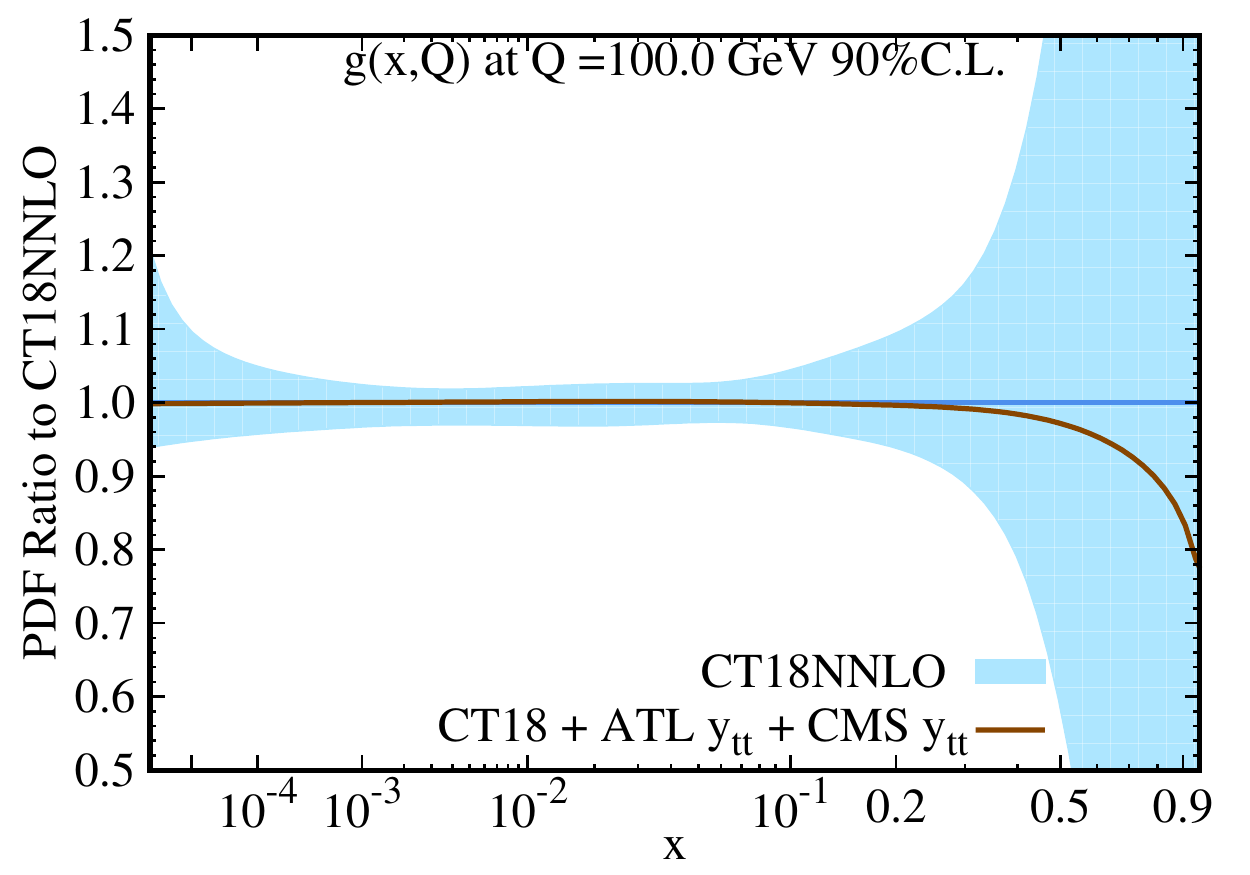}
\caption{{\bf Left}: PDF ratio to CT18NNLO at $Q=2$ GeV for the gluon PDF. {\bf Right}: same as left, but at $Q=100$ GeV. 
The blue band represents the CT18NNLO PDF uncertainty at 90\% CL.}
\label{ttb-fig}        
\end{figure}

Top-quark pair production plays a significant role in setting constraints on PDFs, in particular the gluon, at intermediate to large $x$. 
In global PDF analyses, the impact on PDFs from differential cross section measurements of $t {\bar t}$ production at the LHC complements that from jet production measurements: $t {\bar t}$ and jet production overlap in the $Q$-$x$ plane, but matrix elements and phase-space suppression are different. 
Therefore, constraints on the gluon may be placed at different values of $x$.
On the other hand, challenges in the estimate of full impact from $t {\bar t}$ production on PDFs arise in presence of incompatibility of the 
$t \bar{t}$ measurements with other data sets in the baseline (see related discussion in~\cite{Hou:2019efy}). More realistic estimates account for multiple PDF functional 
forms and some disagreements between the measurements.
In this section, we explore the impact of high-precision $t\bar{t}$ measurements on proton PDFs, 
and illustrate preliminary results of a global QCD analysis in which $t \bar{t}$ single 
differential cross section measurements at $\sqrt{S}=13$ TeV from the ATLAS~\cite{ATLAS:2020ccu} and CMS~\cite{CMS:2018adi} collaborations
are added on top of the CT18 data baseline. The measurements from ATLAS are obtained using events in the all-hadronic channel with 36.1 fb$^{-1}$ of integrated luminosity (IL), while those from CMS are obtained with events containing two leptons with 35.9 fb$^{-1}$ of IL.
We consider the full phase space absolute measurements reported in Table~\ref{toptab}, in which we specify the type of correlated systematic uncertainties published in the experimental papers. We added these measurements one by one to the baseline because statistical correlations were not published for all data sets. 
The measurements from ATLAS and CMS are delivered by using a different binning for the same distribution. The theory predictions for the CMS distributions are obtained by using the \texttt{FastNNLO} grids~\cite{Czakon:2017dip}, that are based on the calculation published in~\cite{Czakon:2015owf}. 
The theory predictions for the ATLAS distributions have been obtained in two steps. 
We have first generated QCD theory predictions at NLO by using in-house \texttt{APPLGrid} fast tables~\cite{Carli:2010rw} from 
the public \texttt{MCFM} computer code~\cite{Campbell:2012uf,Campbell:2015qma}. Then, we calculated bin-by-bin NNLO/NLO $K$-factors using the 
computer code \texttt{MATRIX}~\cite{Grazzini:2017mhc} that is based on the theory predictions calculated in~\cite{Catani:2019hip}.
The mass of the top quark, in the pole mass approximation, has been set as $m_t^{(pole)} = 172.5$ GeV. The factorization and renormalization scales $\mu_F$ and $\mu_R$ respectively, have been chosen according to~\cite{Czakon:2016dgf} where for the $m_{t\bar{t}},p_{T,{t\bar{t}}},y_{t\bar{t}}$, and $y_t$ distributions one uses $\mu_F=\mu_R=H_T/4$ and $H_{T} = \sqrt{m_t^2+p_{T,t}^2} + \sqrt{m_t^2+p_{T,\bar{t}}^2}$\,, for the $p_{T,t}$ distribution one uses $M_{T}=\sqrt{m_t^2+p_{T,t}^2}$, and for the $p_{T,t avg}$ one uses $\mu_F=\mu_R=M_T/2$. EW corrections~\cite{Czakon:2018nun} are also included in this analysis, but their impact on the global fit is negligible.
\begin{table}
\centering
\begin{small}
\begin{tabular}{llll}
\hline
data type & $N_{pt}$ & Exp  & Corr Sys. \\
\hline
$d\sigma/d m_{t\bar{t}}$ & 9 & ATLAS & nuisance par.\\
$d\sigma/d y_{t\bar{t}}$ & 12 & ATLAS & nuisance par.\\
$d\sigma/d H_{T,t\bar{t}}$ & 11 & ATLAS & nuisance par.\\
$d\sigma/d p_{T,t 1}$ & 10 & ATLAS & nuisance par.\\
$d\sigma/d p_{T,t 2}$ & 8 & ATLAS & nuisance par.\\
\hline
\hline
$d\sigma/d m_{t\bar{t}}$ & 7 & CMS & Covariance matr.\\
$d\sigma/d p_{T,t}$ & 6 & CMS & Covariance matr.\\
$d\sigma/d y_{t\bar{t}}$ & 10 & CMS & Covariance matr.\\
$d\sigma/d y_{t}$ & 10 & CMS & Covariance matr.\\
\hline
\end{tabular}
\caption{Details of the single differential distributions for $t \bar{t}$ production considered in our analysis.}
\end{small}
\label{toptab}        
\end{table}
In Figure~\ref{ttb-fig}, we illustrate the pulls on the gluon PDF resulting from a global fit in which the $y_{tt}$ distributions 
from ATLAS and CMS are added on top of the CT18 baseline. The pulls from these distributions from both experiments 
are in the same direction, that is, each of them prefers a softer gluon at large $x$ ($x\gtrsim 0.2$). The individual impact of these measurements on the CT18 
gluon is approximately the same as $\chi^2/N_{pt}$ for both distributions is of order 1. 
A more extensive analysis with complete results will be presented elsewhere~\cite{ttbprep}.

\section{CT18 and the new charm and bottom combination at HERA}
\label{newc-b-HERA}
\begin{figure}[htb!]
\centering
\includegraphics[width=5.4cm,clip]{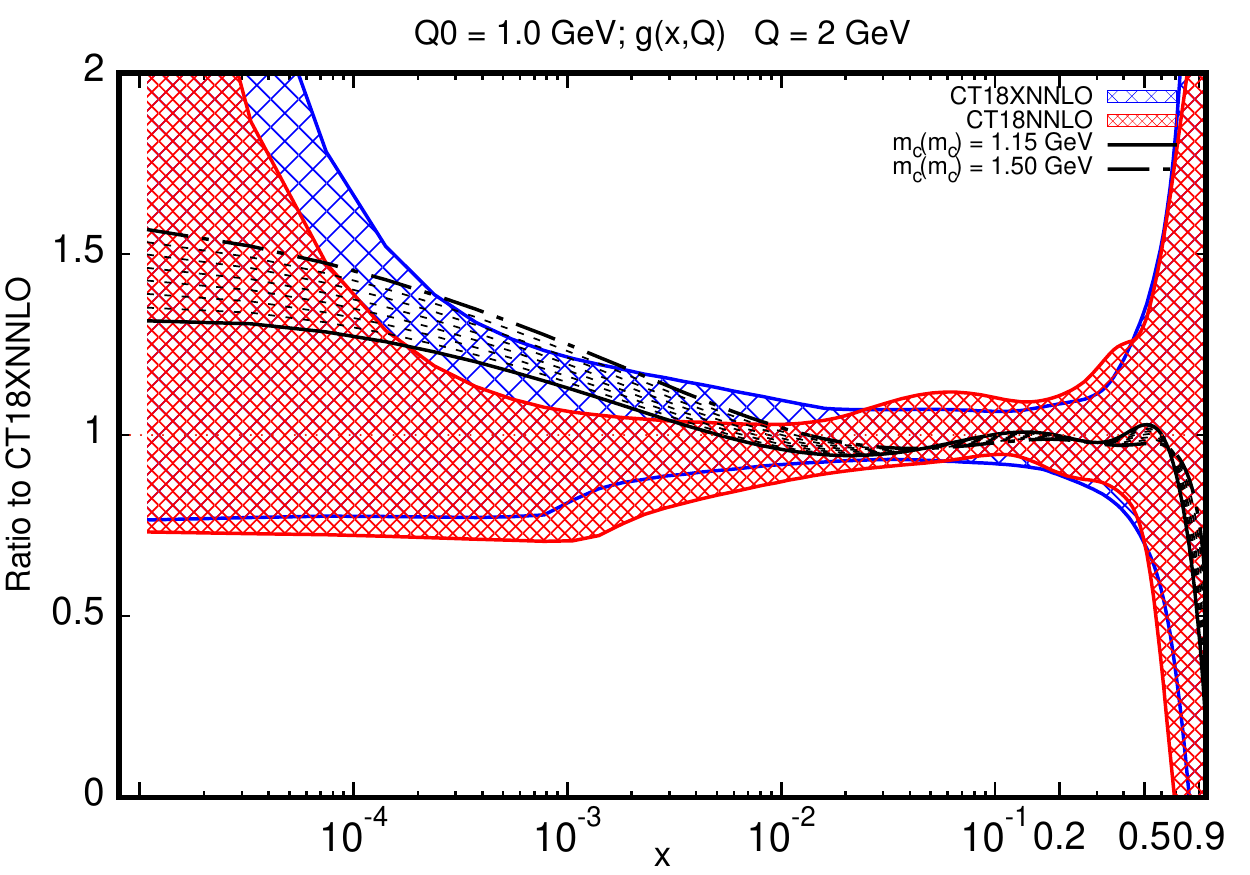}
\includegraphics[width=5.4cm,clip]{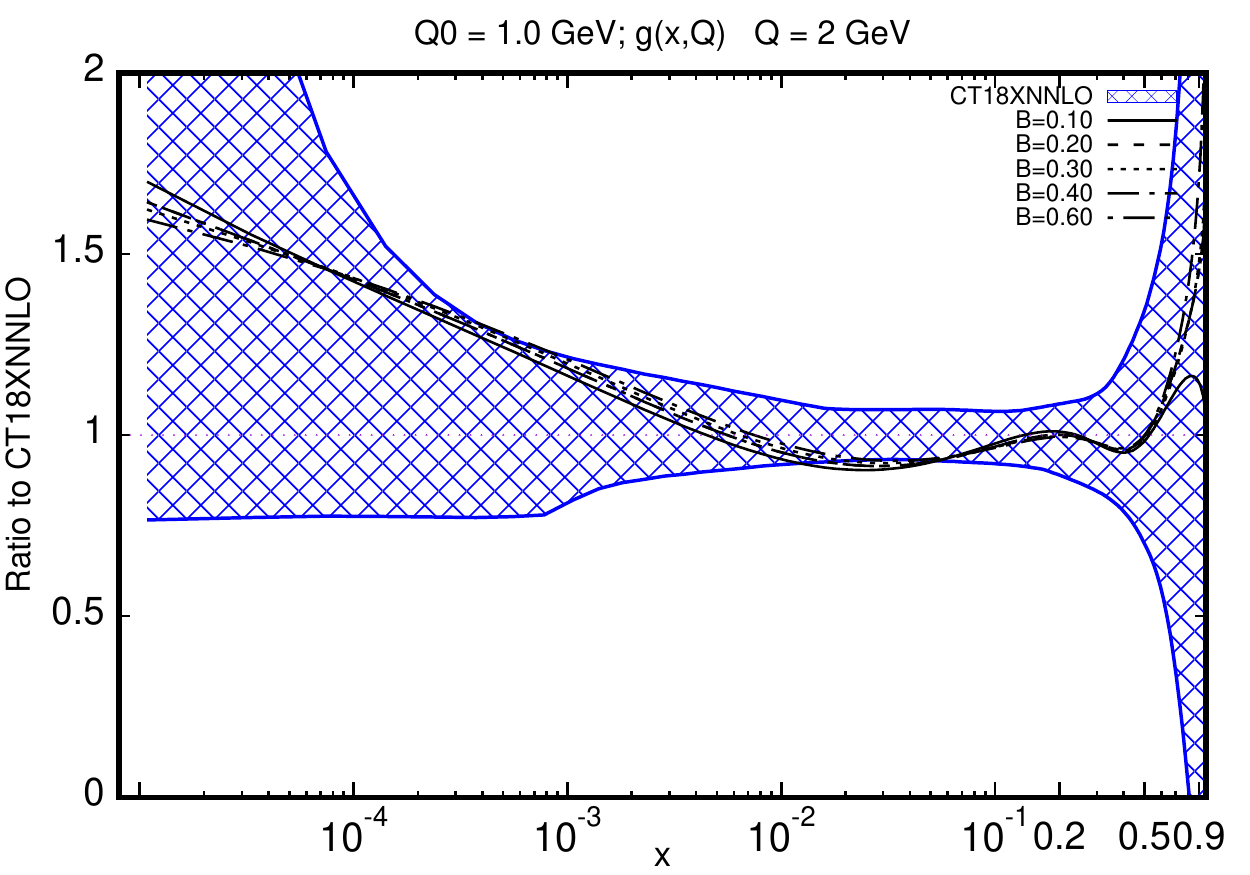}
\caption{Ratio to the CT18XNNLO gluon PDF as a function of $x$ at $Q = 2$ GeV.
{\bf Left}: scan over the $c$-quark mass $\overline{m}_c(m_c)$ while $\overline{m}_b(m_b)$ = 4.18 GeV, and with $Q_0$ = 1 GeV. 
{\bf Right}: scan over the B parameter in $\mu_{DIS}(x)$ with A = 0.5, C = 0.33, and $Q_0$ = 1 GeV. Error bands are shown at 90\% confidence level for
CT18NNLO (red) and CT18XNNLO (blue).} 
\label{hera-fig}        
\end{figure}

To finalize our discussion, in this section we discuss the impact of the most recent $c$- and $b$-quark 
production combined measurements of semi-inclusive
DIS at HERA from the H1 and ZEUS collaborations~\cite{H1:2018flt}, on the CT18 gluon PDF. 
In particular, we illustrate the results of various fits in which these measurements replace their previous version~\cite{H1:2004esl,H1:2012xnw}.
The new measurements exhibit an extended kinematic range of photon virtuality 2.5 GeV$^2$ $\leq Q^2 \leq$ 2000 GeV$^2$ and Bjorken scaling variable 
$3 \cdot 10^{-5} \leq x_\textrm{Bj} \leq 5 \cdot 10^{-2}$, and have reduced uncertainties due to a simultaneous combination of $c$ and $b$ 
cross-section measurements with reduced correlations between them.  
We find that the new $c$ and $b$ combination at HERA~\cite{H1:2018flt} cannot be well described in the CT18 global analysis, and $\chi^2/N_{pt}$ is never below 1.7. For the CT18NNLO fit, we obtained $\chi^2/N_{pt}=1.98$ for $c$ production ($N_{pt}=47$), 
and $\chi^2/N_{pt}=1.25$ for bottom production ($N_{pt}=26$). For the CT18XNNLO fit, instead we obtained $\chi^2/N_{pt}=1.71$ for $c$ and 1.26 for $b$ production.
We observed tensions between these new combined data and several CT18 datasets such as the LHCb 7 and 8 TeV $W/Z$ production data~\cite{LHCb:2015okr,LHCb:2015kwa}, $Z$-rapidity data~\cite{CDF:2010vek} at CDF run-II, CMS 8 TeV single inclusive jet production~\cite{CMS:2016lna}, and $t\bar{t}$ double differential $p_T$ and $y$ cross sections~\cite{CMS:2017iqf}. 
For this reason, these data were not included in the CT18 global analysis~\cite{Hou:2019efy}. 
However, these measurements are very important in PDF determinations because they impose direct constraints on the gluon PDF at intermediate and small $x$.
Therefore, we have re-analyzed these data by performing a series of attempts in which we tried to vary several parameters and settings in the fits.
For example, we performed fits in which we varied the weight assigned to these data. We explored the impact of consistent initial-scale $Q_0$ variations, and tried
different parametrizations for the gluon. We tried to vary the value of the $c$-quark mass $m_c$ in both the pole and $\overline{\textrm{MS}}$ definitions, and we also analyzed the impact of scale variations in which the factorization scale is defined as $\mu_{\textrm{DIS}}(x) = A\sqrt{m_Q^2+B^2/x^C}$ according to saturation models \cite{Golec-Biernat:1998zce,Caola:2009iy}. This is expected to have similar impact to low-$x$ resummation~\cite{Ball:2017otu}. In addition, we explored phase-space suppression in fits in which we varied the S-ACOT-$\chi$ rescaling parameter $\chi=\zeta(1+\zeta^\lambda m_Q^2/Q^2)$. A more detailed description of this study can be found in~\cite{Guzzi:2021gvv}. Here we only provide the results relative to two of the exercises that are described above. In Figure~\ref{hera-fig}, we illustrate the impact on the CT18XNNLO gluon when a scan over the $\overline{\textrm{MS}}$ $\overline{m}_c(m_c)$ is performed (left panel), and the impact from scale variation where parameter $B$ is varied in the $\mu_{\textrm{DIS}}(x)$ scale definition, while A = 0.5 and C = 0.33(right panel). The CT18X fit is a variant of CT18 where $\mu_{F,DIS} = \mu_{\textrm{DIS}}(x)$. Our preliminary findings indicate that the new $c/b$ production measurements at HERA seem to prefer a harder gluon at intermediate and small $x$.
\section{Conclusions}
In this work we have studied HQ production at hadron colliders in recent global QCD analyses and analyzed their impact in collinear proton PDF determinations. 
We discussed S-ACOT-MPS, a new GMVFN scheme which we applied to $pp$ collisions to describe $c/b$ production at central and forward rapidities.
In the near future, it will be technically possible to generate predictions within the S-ACOT-MPS scheme at NNLO with suitable $K$-factors (NNLO/NLO) at hand.
Moreover, it is easy to extend S-ACOT-MPS to other HQ production processes.
We explored the impact of $t\bar{t}$ single differential cross section measurements at the LHC at $\sqrt{S}=$13 TeV on the CT18 PDFs.
The impact of $t\bar{t}$ production at the LHC at $\sqrt{S}=$13 TeV complements that of jet data on the gluon PDF, in particular in the large-$x$ region. 
$t\bar{t}$ and jets overlap in the $Q$-$x$ plane, but being matrix elements and phase-space suppression different, constraints on the gluon PDF 
may be placed at different values of $x$. Overall, the impact is found to be mild with a preference of softer gluon at large $x$. However, this may change when $t\bar{t}$ production in the lepton + jet channel at 13 TeV is included. We analyzed the most recent $c/b$ combination at HERA. We find that these measurements are challenging and 
deserve more attention as they are very important for small-$x$ dynamics. HQ production remains a critical process to constrain correlations between $m_Q, \alpha_s$, and gluon-PDF. 

{\bf Acknowledgments}. The work of M.G. is supported by the NSF under Grant No. PHY-2112025. 
The work of K.X. at is supported in part by the U.S. DOE under Grant No. DE-SC0007914, 
the NSF under Grant No. PHY-1820760, and in part by PITT PACC. The work of P.N. is partially supported by the U.S. DOE under Grant No. DE-SC0010129.
The work of C.P.Y. is partially supported by the NSF under Grant No. PHY-2013791. He is also grateful for the support from the Wu-Ki Tung endowed chair in particle physics.

%
%

\end{document}